\documentclass[11p]{article}

\usepackage{graphicx}
\usepackage{verbatim}
\usepackage{color}
\usepackage{amsmath}
\usepackage{amsfonts}
\usepackage{latexsym}
\usepackage[titles,subfigure]{tocloft}
\usepackage{booktabs}
\usepackage{algorithm}
\usepackage{algorithmic}
\usepackage{amsmath}
\usepackage{lscape}
\usepackage{txfonts}
\usepackage{amssymb}
\usepackage{subfigure}
\usepackage{multicol}
\usepackage{multirow}
\usepackage{chngcntr}
\usepackage{sectsty}
\usepackage[hmargin=1.5in,vmargin=1.5in]{geometry}

\sectionfont{\fontsize{12}{15}\selectfont}

\counterwithout{equation}{section}

\title{Marginally specified models for analyzing multivariate longitudinal binary data}

\author{\"{O}zg\"{u}r Asa$\mbox{r}^{ \ a, *}$ \footnote{ Corresponding author:  \"{O}zg\"{u}r Asar, E-mail address: o.asar@lancaster.ac.uk, Tel.: +44 (0) 1524 593519.}, Ozlem Il$\mbox{k}^{ \ b}$ \\ \\
$^a$ CHICAS, Lancaster Medical School, Lancaster University, UK\\
$^b$ Department of Statistics, Middle East Technical University, Turkey}

\begin{document}

\maketitle

\begin{abstract} 

Marginally specified models have recently become a popular tool for discrete longitudinal data analysis. Nonetheless, they introduce complex constraint equations and model fitting algorithms. Moreover, there is a lack of available software to fit these models. In this paper, we propose a three-level marginally specified model for analysis of multivariate longitudinal binary response data. The implicit function theorem is introduced to approximately solve the marginal constraint equations explicitly. Furthermore, the use of \textit{probit} link enables direct solutions to the convolution equations. We propose an R package \textbf{pnmtrem} to fit the model. A simulation study is conducted to examine the properties of the estimator. We illustrate the model on the Iowa Youth and Families Project data set.

\end{abstract}


\section{Introduction}

Longitudinal data include repeated observations across time which belong to the same subjects/units and which are typically dependent on each other. Often, multiple longitudinal responses, such as multiple health outcomes or distress variables, are of interest. These types of responses involve two types of dependencies: 1) within-response (serial) dependence, and 2) multivariate response dependence at a given time point. Both of these should be taken into account to draw valid statistical inferences, although they might not be of primary interest.    

Traditional longitudinal data models can be threefold: marginal, transition and random effects models (Diggle et al., 2002). Each of these has their pros and cons, and the decision regarding which model to use depends on the scientific interest, i.e., none of them is the best (Gardiner et al., 2009). Recently, marginally specified models (or simply marginalized models) have become popular for discrete longitudinal data analysis which combine the underlying properties of the aforementioned traditional models. Heagerty and Zeger (2000) defined such a model as a re-parameterized version of a model (transition and/or random effects models here) in terms of the marginal mean and additional dependence parameters. The seminal works of Heagerty (1999, 2002) considered likelihood based models by marginalizing the random effects and transition models, respectively. They were two-level logistic regression models that incorporated the marginal covariate effects in the first level, and captured serial dependence in the second level via random effects and transition parameters, respectively. Hence, they permitted multiple inferences at the same time and likelihood based inference for marginal mean parameters. Moreover, in these models marginal regression parameters were shown to be less sensitive to misspecification of dependence structure compared to their traditional counterparts, such as random effects models (Heagerty and Kurland, 2001). The marginalized modeling paradigm were primarily developed for binary longitudinal data (Schildcrout and Heagerty, 2007; Ilk and Daniels, 2007; Lee et al., 2009; along with the aforementioned works of Heagerty), and it has been extended later to ordinal (Caffo and Griswold, 2006; Lee and Daniels, 2007; Lee et al., 2013), count (Lee et al., 2011; Iddi and Molenberghs, 2012) and nominal longitudinal data (Lee and Mercante, 2010). Among these works, Ilk and Daniels (2007) proposed a three-level marginalized model (marginalized transition random effects models, MTREM) for multivariate longitudinal binary data. While the marginal covariate effects were accommodated in the first level, serial and multivariate response dependencies were captured in the second and third levels via transition parameter and random effects, respectively. In this paper, we propose a marginally specified model for multivariate longitudinal binary data by extending MTREM in terms of link functions, i.e., from \textit{logit} to \textit{probit}, and parameter estimation methodology, i.e., from \textit{Bayesian methods} (BM) to \textit{maximum likelihood estimation} (MLE).

{\em probit} and {\em logit} are two widely used link functions in regression analysis of categorical data. While the former is defined as the inverse of the cumulative distribution function (CDF) of standard normal distribution, the latter is defined as the inverse of the CDF of standard logistic distribution. They reflect similar behavior of placing probabilities and are almost indistinguishable except for very low and very high probabilities; {\em logit} lets higher probability in the tails (Hedeker and Gibbons, 2006). Often, large amounts of high quality data are needed to detect substantial differences between the conclusions drawn from the regression models with those link functions (Doksum and Gakso, 1990, cited in Hedeker and Gibbons, 2006, pp. 153). While {\em logit} yields direct interpretation of the regression coefficients, e.g., change in the natural logarithm of the odds ratios, this is more challenging with {\em probit}. However, approximate transitions between {\em logit} and {\em probit} regression parameter estimates are possible (Agresti, 2002; Griswold, 2005). For instance, Johnson et al. (1995, pp. 113-163, cited in Griswold, 2005, pp.85-96) proposed a constant (JKB constant) between these estimates: $\beta_{logit} \cong c*\beta_{probit}$ where $c=(15/16)(\pi/\sqrt{3})$. On the other hand, {\em probit} link usually provides explicit linkage between the levels of marginalized models (Griswold, 2005). However, this is not possible with {\em logit} link. Caffo and Griswold (2006) also discussed the computational advantages of {\em probit} link when it was accompanied with normally distributed components. Multivariate probit modeling literature dates back to the seminal work by Ashford and Sowden (1970). Some recent examples utilizing {\em probit} link in the concept of longitudinal data mixed modeling could be found in Hedeker and Gibbons (2006), Liu and Hedeker (2006), Varin and Czado (2010), Hutmacher and French (2011) among others.  

Semi-parametric methods, namely generalized estimating equations (GEE; Liang and Zeger, 1986), have been widely used for marginal models, especially for discrete response. Nonetheless, they are often inefficient compared to the full likelihood based methods such as MLE and BM. Moreover, a key condition for the consistency of estimates obtained by GEE often fails for transition models (Pepe and Anderson, 1994). BM are also common in longitudinal data literature and have their own properties. Some distinguishing differences are that MLE requires less computational times, and related procedures are more automatized compared to BM (Efron, 1986). In this paper, we consider MLE to avoid the computational burden and possible inconsistencies in a complicated three-level model, for which one of these levels is a transition model. 

Marginally specified models with transition structures introduce marginal constraint equations (Heagerty, 2002; Schildcrout and Heagerty, 2007; Ilk and Daniels, 2007; Lee and Mercante, 2010). Common literature solve these constraints via optimization methods such as Newton-Raphson (N-R) algorithm, which are computationally cumbersome and might yield convergence problems. In this paper, we consider approximately explicit solutions of such constraint equations and propose the use of the \textit{implicit function theorem} for the first time in the scope of marginally specified models. 

There is a lack of available software for analyzing multivariate longitudinal binary data. Limited literature include the works of Shelton et al. (2004), Asar (2012) and Asar and Ilk (2013). Among these, while the former proposed a SAS macro for multivariate longitudinal binary data, the latter proposed two R (R Core Development Team, 2013) packages for multivariate longitudinal data. In this study, we propose an R package \textbf{pnmtrem} for marginalized modeling of multivariate longitudinal binary data. The package is available from the Comprehensive R Archive Network (CRAN) at http://CRAN.R-project.org/package=pnmtrem. Empirical Bayesian estimates of random effects coefficients are also derived, and implementation is included in this package. These estimates allow making subject specific inferences and detecting interesting subjects in the study.  

The paper is organized as follows. In Section 2, we introduce the proposed model and discuss its features. In Section 3, we consider the related parameter estimation procedure. While Section 4 considers a simulation study on the proposed model, Section 5 illustrates its application on a real life data set and discusses the parameter interpretation. We end the paper with the discussion and conclusion part provided in Section 6.
 
\section{Model}

\subsection{General probit normal marginalized transition random effects models, PNMTREM(p)}

Let $Y_{itj}$ be the $j$th ($j=1, \ldots ,k$) response for the $i$th ($i=1, \ldots ,n$) subject at time $t$ ($t=1, \ldots ,T$) and $\boldsymbol X_{itj}$ be the associated set of covariates. $\boldsymbol X_{itj}$ might include time-variant and/or time-invariant covariates. Also let $\Phi(.)$ be the CDF of the standard normal distribution. Use of inverse {\em probit} link yields the following representation of the model:

\begin{eqnarray}
\label{eq:pnmtremg1}
&& P_{itj}^m \equiv P(Y_{itj}=1|\boldsymbol X_{itj})=\Phi(\boldsymbol X_{itj} \boldsymbol \beta ),  \\
\label{eq:pnmtremg2}
&& P_{itj}^t \! \equiv \! P(Y_{itj}=1|y_{i,t-1,j},..,y_{i,t-p,j},
\boldsymbol X_{itj})\!=\!\Phi(\Delta_{itj}+\sum\limits_{m=1}^p \gamma_{itj,m} y_{i,t-m,j}), \\
\label{eq:pnmtremg3}
&& P_{itj}^r \equiv P(Y_{itj}=1|y_{i,t-1,j},...,y_{i,t-p,j},
\boldsymbol X_{itj},b_{it})=\Phi(\Delta_{itj}^*+\lambda_j b_{it}).
\end{eqnarray}

In the first level of the model \eqref{eq:pnmtremg1}, $\boldsymbol \beta$ are the marginal regression coefficients that directly account for the covariate effects on the mean responses, i.e., the covariate effects are not conditioned on either the response history or the random effects. They allow comparing sub-groups of covariates such as females vs. males. Although it is assumed that the intercept and the slopes (covariate effects) are shared by different responses (same $\boldsymbol \beta$ for different responses), the inclusion of response indicator variables as covariates allow different responses to have their own intercepts. Similarly, the inclusion of the interactions of response indicator variables and covariates allow different responses have different slopes. This construction provides model flexibility: it allows one to fit a more parsimonious model when the covariate effects on multiple responses do not differ, which might yield parameter estimates with lower variances. Typical setup of the model assumes that only the covariates at time $t$ have significant effects on the responses at that time point, i.e., $P(Y_{itj}=1|\boldsymbol X_{i1j}, \ldots ,\boldsymbol X_{itj})$ = $P(Y_{itj}=1|\boldsymbol X_{itj})$. Nevertheless, lagged covariates might be included in the model via the design matrix.

In the second level of the model \eqref{eq:pnmtremg2}, the within-subject associations are captured by a Markov model of order $p$. Here the {\em m}th transition parameters, $\gamma_{itj,m}$, can be expressed in terms of covariates. Specifically, $\gamma_{itj,m}= \boldsymbol \alpha_{t,m} \boldsymbol Z_{itj,m}=\alpha_{t1,m} Z_{itj1,m}+...+\alpha_{tl,m} Z_{itjl,m}$ for $m=1,\ldots ,p$, where $\alpha_{tf,m}$ $(f=1, \ldots, l)$ is the time/covariate/order specific transition parameter which accommodates the effect of the past response on the current one by taking into account the interaction between the past response and a subset of covariates as well; $p$ is the order of the transition model and $\boldsymbol Z_{itj}$ are typically a subset of covariates with {\emph l} independent variables. Note that $\boldsymbol Z_{itj}$ have the form of a design matrix, i.e., include 1's on the first column. Choices of $\boldsymbol Z_{itj}$ permit various association structures between the current and past responses. For example, if the effects of the lag-1 responses on the current ones are expected to be different for males and females, then gender could be included in $\boldsymbol Z_{itj,1}$. Similar to the first level, although the transition parameters, $\boldsymbol \alpha_{t,m}$ are shared across multiple responses (common parameters for different responses), the inclusion of interaction(s) between the response indicator variables and the response history allow(s) these parameters to differ for multiple responses.   

In the third level of the model \eqref{eq:pnmtremg3}, the multivariate response dependence and individual variations are accounted by a random effects model. It is possible to observe variations in responses of two subjects even if they have exactly the same  observed covariates and past responses. In such cases, marginal and transition models are inadequate in capturing this subject specific differences. The $b_{it}$ in \eqref{eq:pnmtremg3} measures this unobserved heterogeneity between the subjects at time $t$. $\lambda_j$ is response specific parameter that scales the random effects with respect to response $j$ and accommodates the multivariate response dependence. An approximate correlation among different responses, as a function of $\Delta_{itj}^*$, $\lambda_j$ and $\sigma_{t}^{2}$, can be found in Ilk and Daniels (2007). The $b_{it}$ is subject/time specific random effects coefficient and it is assumed that $b_{it}$ $\sim$ N(0, $\sigma_{t}^{2}$). $b_{it}$ can be rewritten as $b_{it}$=$\sigma_t$ $z_i$ where $z_i$ is a standard normal random variable; this version of $b_{it}$ is useful in numerical integration which will be introduced later. For identifiability, $\lambda_1$ is set to 1. Note that by allowing the random effects to change over time, i.e., by having index $t$ in $b_{it}$, the model accommodates different multivariate response dependencies at different time points.
 
$\Delta_{itj}$ in \eqref{eq:pnmtremg2} is subject/time/response specific intercept that takes the non-linear relationship between the marginal and transition probabilities into account ($P_{itj}^m$ and $P_{itj}^t$, respectively). Similarly, $\Delta_{itj}^*$ in \eqref{eq:pnmtremg3} is the subject/time/response specific intercept that captures the non-linear relationship between the transition and random effects probabilities ($P_{itj}^t$ and $P_{itj}^r$, respectively). 

This three-level model specification of PNMTREM completes the multivariate distribution of the multivariate longitudinal binary data. One of the inherited features of PNMTREM from the original setup of MTREM is that the conditional mean of the responses given all set of covariates is equal to the conditional mean of the responses given the covariate history, i.e., $E(Y_{itj}|\boldsymbol X_{iqj},q=1, \ldots, T)=E(Y_{itj}|\boldsymbol X_{isj}, s \leq t)$. This assumption is vital for the validity of the marginal constraint equation which will be introduced later while linking the levels of the model. However, the assumption is meaningful for exogenous covariates (covariates which do not depend on response history at time $t$) but not meaningful for the endogenous ones (covariates which depend on response history at time $t$). 

\subsection{First order probit normal marginalized transition random effects models, PNMTREM(1)}

Here, we discuss first order model, PNMTREM(1), which is a specialized form of the general model, PNMTREM(p). PNMTREM(1) considers only the effects of lag-1 responses on the current ones in the second level of the model formulation and the related modeling framework is given by

\begin{eqnarray}
\label{eq:pnmtrem11}
&& P_{itj}^m \equiv P(Y_{itj}=1|\boldsymbol X_{itj})=\Phi(\boldsymbol X_{itj} \boldsymbol \beta),  \\
\label{eq:pnmtrem12}
&& P_{itj}^t \equiv P(Y_{itj}=1|y_{i,t-1,j},\boldsymbol X_{itj})\!=\!\Phi(\Delta_{itj}+\gamma_{itj,1} y_{i,t-1,j}), \\
\label{eq:pnmtrem13}
&& P_{itj}^r \equiv P(Y_{itj}=1|y_{i,t-1,j},\boldsymbol X_{itj},b_{it})=\Phi(\Delta_{itj}^*+\lambda_j b_{it}),
\end{eqnarray} 

\noindent where $b_{it}$ $\sim$ N(0, $\sigma_{t}^{2}$) and $b_{it}$=$z_i$ $\sigma_t$, $z_i$ $\sim$ N(0,1); $\lambda_1$=1. Again, $\gamma_{itj,1}=\boldsymbol \alpha_{t,1} \boldsymbol Z_{itj,1}=\alpha_{t1,1} Z_{itj1,1}+...+\alpha_{tl,1} Z_{itjl,1}$ where $\boldsymbol Z_{itj,1}$ are a subset of covariates. Note that throughout we call this model as the $t \geq 2$ model.

Since for baseline $(t=1)$ no history data are available, the second level of PNMTREM(1) is not valid anymore. Additionally, it is common in longitudinal studies that baseline data reflect more or less variability and have different covariate effects compared to later time points. In the light of these arguments, a separate model is constructed for $t=1$. The baseline model is given by

\begin{eqnarray}
\label{eq:pnmtremb1}
&& P_{i1j}^m \equiv P(Y_{i1j}=1|\boldsymbol X_{i1j})=\Phi(\boldsymbol X_{i1j} \boldsymbol \beta^*),  \\
\label{eq:pnmtremb2}
&& P_{i1j}^r \equiv P(Y_{i1j}=1|\boldsymbol X_{i1j},b_{i1})=\Phi(\Delta_{i1j}^*+\lambda_j^* b_{i1}).
\end{eqnarray}

Here, $b_{i1}$ $\sim$ N(0, $\sigma_{1}^{2}$) and $b_{i1}$=$z_i$ $\sigma_1$, $z_i$ $\sim$ N(0,1); $\lambda_1^*$=1. Note that throughout we call this model as the {\em baseline} model. \\

\subsubsection{Linking levels of PNMTREM(1) for t $\geq$ 2 model}

To be a valid probabilistic model, the levels of PNMTREM(1) are connected to each other by a set of constraint equations.\\

\noindent \textbf{Linking first and second levels of PNMTREM(1)} \\

Level 1 \eqref{eq:pnmtrem11} and level 2 \eqref{eq:pnmtrem12} of PNMTREM(1) are linked via the marginal constraint equation,

\begin{equation}
\label{eq:margconst1}
P(Y_{itj}=1|\boldsymbol X_{itj})= \sum\limits_{y_{i,t-1,j}} P(Y_{itj}=1|y_{i,t-1,j},\boldsymbol X_{itj}) P(y_{i,t-1,j}|\boldsymbol X_{i,t-1,j}),
\end{equation}

\noindent which is equivalent to

\begin{equation}
\label{eq:margconst2.5}
\Phi(\boldsymbol X_{itj} \boldsymbol \beta) = \sum\limits_{y_{i,t-1,j}=0}^1 \Phi(\Delta_{itj}+\gamma_{itj,1} y_{i,t-1,j}) (\Phi(\boldsymbol X_{i,t-1,j} \boldsymbol \beta))^{y_{i,t-1,j}} (1-\Phi(\boldsymbol X_{i,t-1,j} \boldsymbol \beta))^{(1-y_{i,t-1,j})},
\end{equation}

\noindent or, in a simpler form, equivalent to,

\begin{equation}
\label{eq:margconst3}
\Phi(\boldsymbol X_{itj} \boldsymbol \beta) = \Phi(\Delta_{itj}) (1-\Phi(\boldsymbol X_{i,t-1,j} \boldsymbol \beta)) + \Phi( \Delta_{itj}+ \gamma_{itj,1}) \Phi(\boldsymbol X_{i,t-1,j} \boldsymbol \beta).
\end{equation}

Note that for $t=2$, $\boldsymbol \beta^*$ replace $\boldsymbol \beta$ as the multiplier of lag-1 covariates in the marginal constraint equation and yields

\begin{equation}
\label{eq:margconst4}
\Phi(\boldsymbol X_{i2j} \boldsymbol \beta) = \Phi(\Delta_{i2j}) (1-\Phi(\boldsymbol X_{i,1,j} \boldsymbol \beta^*)) + \Phi( \Delta_{i2j}+ \gamma_{i2j,1}) \Phi(\boldsymbol X_{i,1,j} \boldsymbol \beta^*).
\end{equation}

Although the rest of the discussion will be based on \eqref{eq:margconst3}, we take the difference into account when necessary. The non-linear equation given in \eqref{eq:margconst3} does not permit writing $\Delta_{itj}$ in terms of $\boldsymbol \beta$ and $\gamma_{itj,1}$ (or $\boldsymbol \alpha_{t,1}$), explicitly. Luckily, the implicit function theorem (IFT; Krantz and Parks, 2003) allows us finding an explicit solution of \eqref{eq:margconst3}, though an approximate one, for $\Delta_{itj}$ in terms of $\boldsymbol \beta$ and $\boldsymbol \alpha_{t,1}$.\\

\noindent \textbf{Application of IFT to PNMTREM(1)}\\

Let F be a function of $\boldsymbol X_{itj}$, $\boldsymbol X_{it-1j}$, $\boldsymbol \beta$, $\Delta_{itj}$, $\boldsymbol \alpha_{t,1}$ and $\boldsymbol Z_{itj,1}$ such that (by rewriting \eqref{eq:margconst3})
\small
\begin{equation}
\label{eq:Fift}
F(\boldsymbol X_{itj}, \boldsymbol X_{it-1j}, \boldsymbol \beta, \Delta_{itj}, \boldsymbol \alpha_{t,1}, \boldsymbol Z_{itj,1})= \\
\Phi(\boldsymbol X_{itj} \boldsymbol \beta) - \Phi(\Delta_{itj}) (1-\Phi(\boldsymbol X_{i,t-1,j} \boldsymbol \beta)) - \Phi(\Delta_{itj}+\boldsymbol \alpha_{t,1} \boldsymbol Z_{itj,1}) \Phi(\boldsymbol X_{i,t-1,j} \boldsymbol \beta)=0. 
\end{equation}
\normalsize
Then, by IFT and first order implicit differentiation (first order approximation), $\Delta_{itj}$ could be obtained by 

\begin{equation}
\label{eq:deltaexpl}
\Delta_{itj} = - \frac{\frac{\partial F}{\partial \boldsymbol \beta} \Big |_{(\boldsymbol \beta_0,\boldsymbol \alpha_{t,10},\Delta_{itj0})}} {\frac{\partial F}{\partial \Delta_{itj}} \Big |_{(\boldsymbol \beta_0,\boldsymbol \alpha_{t,10},\Delta_{itj0})}} (\boldsymbol \beta- \boldsymbol \beta_0) -\frac{\frac{\partial F}{\partial \boldsymbol \alpha_{t,1}} \Big |_{(\boldsymbol \beta_0,\boldsymbol \alpha_{t,10},\Delta_{itj0})}} {\frac{\partial F}{\partial \Delta_{itj}} \Big |_{(\boldsymbol \beta_0,\boldsymbol \alpha_{t,10},\Delta_{itj0})}} (\boldsymbol \alpha_{t,1}- \boldsymbol \alpha_{t,10}), 
\end{equation}

\noindent where 

\begin{align}
\label{eq:fpartial1}
\frac{\partial F}{\partial \boldsymbol \beta} &=\boldsymbol X_{itj} \phi(\boldsymbol X_{itj} \boldsymbol \beta) + \Phi(\Delta_{itj})(\phi(\boldsymbol X_{i,t-1,j}\boldsymbol \beta)) \boldsymbol X_{i,t-1,j} 
- \Phi(\Delta_{itj}+ \boldsymbol \alpha_{t,1} \boldsymbol Z_{itj,1}) \phi(\boldsymbol X_{i,t-1,j} \boldsymbol \beta) \boldsymbol X_{i,t-1,j}, \nonumber \\
\frac{\partial F}{\partial \Delta_{itj}} &= - \phi(\Delta_{itj})(1-\Phi(\boldsymbol X_{i,t-1,j} \boldsymbol \beta)) - \phi(\Delta_{itj}+ \boldsymbol \alpha_{t,1} \boldsymbol Z_{itj,1})(\Phi(\boldsymbol X_{i,t-1,j} \boldsymbol \beta)), \nonumber\\
\frac{\partial F}{\partial \boldsymbol \alpha_{t,1}} &= - \phi(\Delta_{itj}+ \boldsymbol \alpha_{t,1} \boldsymbol Z_{itj}) \Phi(\boldsymbol X_{i,t-1,j} \boldsymbol \beta) \boldsymbol Z_{itj,1} .
\end{align}

Here, $\phi(.)$ is the probability density function of the standard normal distribution and $\boldsymbol \beta_{0}, \boldsymbol \alpha_{t,10}$ and $\Delta_{itj0}$ are the components of $\boldsymbol P_0$ around which IFT searches for solution. For $t=2$, $\boldsymbol \beta^*$ replace $\boldsymbol \beta$ as the multiplier of lag-1 covariates.

From \eqref{eq:deltaexpl} and \eqref{eq:fpartial1}, it can be seen that $\Delta_{itj}$ is explicit and deterministic function of $\boldsymbol X_{itj}$, $\boldsymbol X_{it-1j}$, $\boldsymbol \beta$, $\boldsymbol \alpha_{t,1}$ and $\boldsymbol Z_{itj,1}$, i.e., $\Delta_{itj}$=$\Delta_{itj}(\boldsymbol X_{itj}, \boldsymbol X_{it-1j}, \boldsymbol \beta, \boldsymbol \alpha_{t,1}, \boldsymbol Z_{itj,1})$. Here, we shall note that $\Delta_{i2j}$ is function of both $\boldsymbol \beta$ and $\boldsymbol \beta^*$. The $\boldsymbol \beta_0$ and $\boldsymbol \alpha_{t,10}$ components of $\boldsymbol P_0$ for PNMTREM are taken to be $\boldsymbol 0$, since the hypothesis tests about the significances of $\boldsymbol \beta$ and $\boldsymbol \alpha_{t,1}$ place null hypotheses which assume the equality of those parameters to be $\boldsymbol 0$. $\Delta_{itj0}$ is obtained by solving \eqref{eq:Fift} when $\boldsymbol \beta_0$ and $\boldsymbol \alpha_{t,10}$ are equal to $\boldsymbol 0$. This yields $\Delta_{itj0}=0$ when $t > 2$. We only employ N-R to obtain $\Delta_{i2j0}$. This has very fast convergence based on our experience, due to the simple form of the related function, given in \eqref{eq:Fift}.\\

\noindent \textbf{Linking second and third levels of PNMTREM(1)}\\

Level 2 \eqref{eq:pnmtrem12} and level 3 \eqref{eq:pnmtrem13} of PNMTREM(1) are linked via a convolution equation given by

\begin{equation}
\label{eq:convolution1}
P(Y_{itj}=1|y_{i,t-1,j},\boldsymbol X_{itj})= \int P(Y_{itj}=1|y_{i,t-1,j},\boldsymbol X_{itj},b_{it})dF(b_{it}), 
\end{equation}

\noindent which is equivalent to

\begin{equation}
\label{eq:convolution3}
\Phi(\Delta_{itj}+ \boldsymbol \alpha_{t,1} \boldsymbol Z_{itj,1} y_{i,t-1,j}) = \int \Phi(\Delta_{itj}^*+\lambda_j b_{it}) f(b_{it}) db_{it}.
\end{equation}

Following Griswold (2005), we can obtain 

\begin{equation}
\label{eq:deltastarexpl}
\Delta_{itj}^*=\sqrt{1+\lambda_j^2 \sigma_t^2} \; (\Delta_{itj}+ \boldsymbol \alpha_{t,1} \boldsymbol Z_{itj,1} y_{i,t-1,j}).
\end{equation}

Related proof can be found in Appendix A. From \eqref{eq:deltastarexpl}, it can be seen that $\Delta_{itj}^*$ is explicit and deterministic function of $\Delta_{itj}$ (hence, $\Delta_{itj}^*$ is function of $\boldsymbol X_{itj}$, $\boldsymbol X_{i,t-1,j}$ and $\boldsymbol \beta$), $\boldsymbol \alpha_{t,1}$ $\boldsymbol Z_{itj,1}$, $y_{it-1j}$, $\lambda_j$ and $\sigma_{t}$, i.e., $\Delta_{itj}^*=\Delta_{itj}^*(\boldsymbol X_{itj}, \boldsymbol X_{i,t-1,j}, \boldsymbol \beta, \Delta_{itj}, \boldsymbol \alpha_{t,1}, \boldsymbol Z_{itj,1}, y_{it-1j}, \lambda_j, \sigma_{t})$.

\subsubsection{Linking levels of PNMTREM(1) for the baseline model}

The levels of the baseline model are linked to each other via the following convolution equation:

\begin{equation}
\label{eq:convolution1b}
P(Y_{i1j}=1|\boldsymbol X_{i1j})= \int P(Y_{i1j}=1|\boldsymbol X_{i1j},b_{i1})dF(b_{i1}). 
\end{equation}

Again, following Griswold (2005), we can obtain $\Delta_{i1j}^*$ as an explicit function of $\boldsymbol X_{i1j}$, $\boldsymbol \beta^*$, $\lambda^*_j$ and $\sigma_1$ such that

\begin{equation}
\label{eq:deltastar1expl}
\Delta_{i1j}^* = \sqrt{1+{\lambda_{j}^*}^{2} \sigma_1^2} \; \boldsymbol X_{i1j} \boldsymbol \beta^*.
\end{equation} 

Related proof is very similar to the one for $t \geq 2$ model and can be easily adapted from it. 

Unlike {\em logit} link, the use of {\em probit} link in MTREM allows us directly writing the levels in terms of each other. This allows us maximizing the likelihood and obtaining the maximum likelihood estimates (MLE) of the parameters without taking the derivatives of $\Delta_{i1j}^*$, $\Delta_{itj}$ and $\Delta_{itj}^*$ $(t \geq 2)$ with respect to former level parameters. This eases the related MLE derivations and decreases computational time. We will discuss these aspects later.
 
\section{Estimation}

\subsection{Likelihood function of PNMTREM(1)}
 
We assume two different models for a given multivariate longitudinal binary data set in the PNMTREM framework: baseline and $t \geq 2$ models. The related likelihood function of PNMTREM(1) is the product of two likelihood functions belonging to these models. Rewriting the random effects $b_{i1}$ and $b_{it}$ as $b_{i1}=\sigma_1 z_i$ and $b_{it}=\sigma_t z_i$ yields this likelihood to be

\begin{equation}
\label{eq:mtremlik2}
L(\boldsymbol \theta|\boldsymbol y) = L_1(\boldsymbol \theta_1|\boldsymbol y_1) L_2(\boldsymbol \theta_2|\boldsymbol y_2),
\end{equation}
where
\begin{align}
\label{eq:baselinelik}
L_1(\boldsymbol \theta_1|\boldsymbol y_1) &=\prod_{i=1}^{N} \int \prod_{j=1}^{k} \left(P^{r}_{i1j}\right)^{y_{i1j}} \left(1-P^{r}_{i1j}\right)^{1-y_{i1j}}\phi(z_i)dz_i,
\end{align}
\vspace{-0.2in}
\begin{align}
\label{eq:mainlik}
L_2(\boldsymbol \theta_2|\boldsymbol y_2) &=\prod_{i=1}^{N} \prod_{t=2}^{T} \int \prod_{j=1}^{k} \left(P^{r}_{itj}\right)^{y_{itj}} \left(1-P^{r}_{itj}\right)^{1-y_{itj}}\phi(z_i)dz_i.
\end{align}

Here, $\boldsymbol \theta=(\boldsymbol \theta_1, \boldsymbol \theta_2)$, where $\boldsymbol \theta_1=(\boldsymbol \beta^*, \boldsymbol \lambda^*, \sigma_1^2)$ with $\boldsymbol \lambda^*=(\lambda_2^*, \ldots, \lambda_k^*)$ and $\boldsymbol \theta_2=(\boldsymbol \beta, \boldsymbol \alpha_{t,1},\boldsymbol \lambda, \boldsymbol \sigma^2)$ with $\boldsymbol \lambda=(\lambda_2, \ldots, \lambda_k)$ and $\boldsymbol \sigma^2=(\sigma_2^2, \ldots, \sigma_T^2)$, are parameter vectors for baseline and $t \geq 2$ models, respectively; $\boldsymbol y_1$ and $\boldsymbol y_2$ are the observed response matrices at baseline and $t \geq 2$ time points, respectively. Although these two likelihoods seem to be independent, they are connected to each other via the estimates of $\boldsymbol \beta^*$, i.e., $\hat{\boldsymbol \beta^*}$, for $t=2$ due to the marginal constraint equation (see \eqref{eq:margconst4}). We consider the estimation of $log(\sigma_t)$ for $t=1,\ldots,T$, instead of directly estimating  $\sigma_t$ or $\sigma_t^2$, due to computational aspects, since taking logarithm of the variance components extends the related parameter space from the interval of $[0,+\infty)$ to $(-\infty,+\infty)$. Turning back to the estimates of $\sigma_t$ or $\sigma_t^2$ is possible by the invariance property of maximum likelihood estimates (MLE), and the related variance estimates could be obtained by the delta method.

Maximizing the likelihood function given in \eqref{eq:mtremlik2} needs numerical methods while taking one-dimensional integrals over the standard normal distribution. It is well known that for approximating one-dimensional integrals, e.g., for constant random effects over time or independent random effects over time, Gauss-Hermite quadrature is a successful method (Agresti, 2002; McCulloch et al., 2008). It is reported that a 20-point Gauss-Hermite quadrature is usually enough to achieve an accurate approximation for likelihood functions (McCulloch et al., 2008, pp. 329; Lesaffre and Spiessens, 2001). Note that Ilk and Daniels (2007) considered BM, specifically Markov Chain Monte Carlo methods, for MTREM as the parameter estimation methodology, which did not require taking integrals over the random effects distribution in likelihoods. However, this method yielded the parameter estimation process taking long time. 

First partial derivatives of the log-likelihood functions of PNMTREM do not permit obtaining explicit solutions to the MLE of the parameters. Therefore, optimization techniques are needed. N-R requires the calculation of first and second partial derivatives of the log-likelihood functions. However, for PNMTREM(1) even the first partial derivatives of the log-likelihoods have very complex forms, hence the use of N-R is not appropriate. Luckily, Fisher-Scoring Algorithm (F-S) solves the log-likelihood functions by using only the first partial derivatives (Hedeker and Gibbons, 2006, pp. 162-165). Another great feature of F-S is that the inverse of the expected information matrix at convergence is a consistent estimator of the large sample variance-covariance matrix of the model parameters.

\subsection{Maximum likelihood estimation of the baseline parameters ($\boldsymbol \theta_1$)}

Maximizing the log-likelihood function of the baseline model, $L_1(\boldsymbol \theta_1|\boldsymbol y_1)$, with respect to $\boldsymbol \theta_1$ yields
\small
\begin{align}
\label{eq:apploglik1}
\frac{\partial \mbox{log} \left(L_1(\boldsymbol \theta_1|\boldsymbol y_1)\right)}{\partial \boldsymbol \theta_1} \approx \sum_{i=1}^{N} \frac {1}{h(Y_{i1}|\boldsymbol \theta_1)} \frac{\partial h(Y_{i1}|\boldsymbol \theta_1)}{\partial \boldsymbol \theta_1},
\end{align}
\normalsize
\noindent where
\begin{align}
\label{eq:apph1}
h(Y_{i1}|\boldsymbol \theta_1) \approx \sum_{q=1}^{20} w_q \; \underbrace{ exp \left[ \sum_{j=1}^k \left( Y_{i1j} \mbox{log} \left(\Phi(d_{i1jq})\right) + (1-Y_{i1j}) \mbox{log}\left(1-\Phi(d_{i1jq})\right) \right)\right]}_{\ell(Y_{i1}|\boldsymbol \theta_1)},
\end{align} 
\vspace{-0.15in}
\begin{align}
\label{eq:apphd1}
\frac{\partial h(Y_{i1}|\boldsymbol \theta_1)}{\partial \boldsymbol \theta_1} \approx \sum_{q=1}^{20} w_q \; \left\{  \ell(Y_{i1}|\boldsymbol \theta_1) \left\{ \sum_{j=1}^{k} \left[ \frac{\partial d_{i1jq} }{\partial \boldsymbol \theta_1} \phi(d_{i1jq}) \left( \frac{Y_{i1j}-\Phi(d_{i1jq})}{\left(\Phi(d_{i1jq}) \right) \left( 1-\Phi(d_{i1jq})\right)} \right) \right] \right\}\right\},
\end{align}
\vspace{-0.15in}
\begin{align}
\label{eq:appd1}
d_{i1jq}=\sqrt{1+{\lambda_j^*}^2 e^{2c_1}} \; (\boldsymbol X_{i1j} \boldsymbol \beta^*)+\lambda_j^*e^{c_1}\sqrt{2}\;z_q.
\end{align}

Here $log(\sigma_1)$ is equated to $c_1$ for simplicity of notation and $(z_q, w_q)$ for $q=1,\ldots,20$ are Gauss-Hermite quadrature points and weights, respectively which are available in Abramowitz and Stegun (1972). Details of $\frac{\partial d_{i1jq} }{\partial \boldsymbol \theta_1}$ can be found in Appendix B.1.

\subsection{Maximum likelihood estimation of the $t \geq 2$ parameters ($\boldsymbol \theta_2$)}

Similar to the baseline model, maximizing the log-likelihood function of the $t \geq 2$ model with respect to $\boldsymbol \theta_2$ yields

\small
\begin{align}
\label{eq:apploglik2}
\frac{\partial log\left(L_2(\boldsymbol \theta_2|\boldsymbol y_2)\right)}{\partial \boldsymbol \theta_2} \approx \sum_{i=1}^{N} \sum_{t=2}^{T} \frac {1}{h(Y_{it}|\boldsymbol \theta_2)} \frac{\partial h(Y_{it}|\boldsymbol \theta_2)}{\partial \boldsymbol \theta_2},
\end{align}
\normalsize
\noindent where
\begin{align}
\label{eq:apph2}
h(Y_{it}|\boldsymbol \theta_2) \approx \sum_{q=1}^{20} w_q \; \underbrace{exp \left[ \sum_{j=1}^k \left( Y_{itj}log\left(\Phi(d_{itjq})\right) + (1-Y_{itj}) log\left(1-\Phi(d_{itjq})\right) \right)\right]}_{\ell(Y_{it}|\boldsymbol \theta_2)},
\end{align}
\vspace{-0.15in}
\begin{align}
\label{eq:apphd2}
\frac{\partial h(Y_{it}|\boldsymbol \theta_2)}{\partial \boldsymbol \theta_2} \approx \sum_{q=1}^{20} w_q \; \left\{  \ell(Y_{it}|\boldsymbol \theta_2) \left\{ \sum_{j=1}^{k} \left[ \frac{\partial d_{itjq} }{\partial \boldsymbol \theta_2} \phi(d_{itjq}) \left( \frac{Y_{itj}-\Phi(d_{itjq})}{\left(\Phi(d_{itjq}) \right) \left( 1-\Phi(d_{itjq})\right)} \right) \right] \right\}\right\},
\end{align}
\vspace{-0.15in}
\begin{align}
\label{eq:appd2}
d_{itjq}=\sqrt{1+\lambda_j^2 e^{2c_t}}\left(\Delta_{itj}+\boldsymbol \alpha_{t,1} \boldsymbol Z_{itj} y_{it-1j}\right) +\lambda_je^{c_t} \sqrt{2} z_q.
\end{align}

Here $c_t=log(\sigma_t)$ for $t \geq 2$, and $(z_q,w_q)$ for $q=1,\ldots,20$ are Gauss-Hermite quadrature points and weights. Also note that explicit solution of $\Delta_{itj}$ is given in \eqref{eq:deltaexpl}. Details of $\frac{\partial d_{itjq} }{\partial \boldsymbol \theta_2}$ can be found in Appendix B.2.

\subsection{Application of Fisher-Scoring algorithm}

As stated earlier, the MLEs of the parameters are obtained iteratively by Fisher-Scoring Algorithm (F-S) and the related algorithm is given by
\begin{align}
\label{eq:fst1}
\boldsymbol \theta_s^{(m+1)}=\boldsymbol \theta_s^{m}+I(\boldsymbol \theta_s^m)^{-1} \frac{\partial log\left(L_s(\boldsymbol \theta_s^m|\boldsymbol y_s)\right)}{\partial \boldsymbol \theta_s^m},
\end{align}

\noindent where $\boldsymbol s=(1,2)$; $s=1$ corresponds to the baseline model and $s=2$ corresponds to the $t \geq 2$ model; $m$ represents the F-S step and $I(\boldsymbol \theta_s)$ is an empirical and consistent estimator of the information matrix. $I(\boldsymbol \theta_s)$ can be calculated by

\begin{align}
\label{inft1}
I(\boldsymbol \theta_1)=\sum_{i=1}^{N} h(Y_{i1j}|\boldsymbol \theta_1)^{-2} \left( \frac{\partial h(Y_{i1j}|\boldsymbol \theta_1)}{\partial \boldsymbol \theta_1}\right) \left( \frac{\partial h(Y_{i1j}|\boldsymbol \theta_1)}{\partial \boldsymbol \theta_1}\right)^T
\end{align}
\noindent and
\begin{align}
\label{inftgeq2}
I(\boldsymbol \theta_2)=\sum_{i=1}^{N} \left(\sum_{t=2}^{T} \frac {1} {h(Y_{itj}|\boldsymbol \theta_2)} \frac{\partial h(Y_{itj}|\boldsymbol \theta_2)}{\partial \boldsymbol \theta_2} \right) \left(\sum_{t=2}^{T} \frac {1} {h(Y_{itj}|\boldsymbol \theta_2)} \frac{\partial h(Y_{itj}|\boldsymbol \theta_2)}{\partial \boldsymbol \theta_2} \right)^{T}.
\end{align}

Since $c_1$ is time specific and $\lambda^*_j$ is response specific for baseline and $c_t$ and $\boldsymbol \alpha_{t,1}$ are time specific and $\lambda_j$ is response specific for $t \geq 2$, the forms of $I(\boldsymbol \theta_1)$ and $I(\boldsymbol \theta_2)$ are quite different compared to the ones for $\boldsymbol \beta^*$ and $\boldsymbol \beta$ for baseline and $t \geq 2$ models, respectively. Details can be found in the supplementary material to this article.

\subsection{Empirical Bayesian estimation of random effects coefficients}

To calculate the individual probabilities such as $P(Y_{i1j}=1|\boldsymbol X_{i1j},b_{i1})$ and $P(Y_{itj}=1|\boldsymbol X_{itj},y_{it-1j},b_{it})$ for $t \geq 2$, we need the estimates of $\Delta_{i1j}^*$, $\lambda_j^*$, $b_{i1}=\sigma_1 z_i$ for the baseline model and $\Delta_{itj}^*$, $\lambda_j$, $b_{it}=\sigma_t z_i$ for the $t \geq 2$ model.

Given the MLEs of $\boldsymbol \theta_1=(\boldsymbol \beta^*, \boldsymbol \lambda^*, c_1=log(\sigma_1))$ and $\boldsymbol \theta_2=(\boldsymbol \beta, \boldsymbol \alpha_{t,1}, \boldsymbol \lambda, \boldsymbol c=log(\boldsymbol \sigma))$, we can obtain the Empirical Bayes estimators of $b_{it}$, $\tilde{b}_{it}$ ($t=1,\ldots,T$) by solving the posterior score equations of $z_i$ (Heagerty, 1999). The posterior distribution of $z_i$ is proportional to the conditional distribution of the observed data given $z_i$, $[Y_i|z_i]$, times the prior distribution of $z_i$, and $\hat{z}_i$ can be obtained as the mode of log-posterior distribution. This requires equating the first partial derivative of the natural logarithm of the posterior distribution of $z_i$ with respect to $z_i$ to 0 and then solving the score equations for $z_i$. The related score equation is given by

\begin{align}
\label{eq:empbay}
\left\{ \sum_{t=1}^{T} \sum_{j=1}^{k} \frac{\hat{\lambda}_j \hat{\sigma}_{t} \phi(\hat{d}_{itj})\left(Y_{itj}-\Phi(\hat{d}_{itj})\right)}{\Phi(\hat{d}_{itj})\left(1-\Phi(\hat{d}_{itj}) \right)} \right\}-z_i=0,
\end{align}

\noindent where $\hat{d}_{itj}=\hat{\Delta}_{itj}^{*}+\hat{\lambda}_j \hat{\sigma}_{t} z_i$ and $\hat{\Delta}_{itj}^{*}$ are obtained by using the MLEs of $\boldsymbol \theta_1$ and $\boldsymbol \theta_2$. Since \eqref{eq:empbay} does not permit closed solutions for $z_i$, N-R algorithm is utilized. 
 
\section{Simulation study}

We conducted a Monte Carlo simulation study to examine the bias and variance of the marginal mean parameters. In each replications, we simulated data sets under PNMTREM(1) which included bivariate binary responses and two associated covariates for 250 subjects with 4 follow-ups. We considered different sets of covariates for baseline and $t \geq 2$ time points. Moreover, we considered varying effects of the covariates for these time points, i.e., $\boldsymbol \beta^* \nequiv \boldsymbol \beta$. 

For $t=1$, we considered true parameter settings of $\boldsymbol \beta^*=(\beta_0^*, \beta_1^*) = (-1, 1.9)$, $\boldsymbol \lambda^*=(\lambda_1^*, \lambda_2^*)=(1, 1.07)$ and $b_{i1} \sim N(0,\sigma_1^2)$, $\sigma_1=0.7$. $X_1$ was generated from $Uniform(0,1)$. On the other hand, for $t \geq 2$, we considered parameter settings of $\boldsymbol \beta=(\beta_0, \beta_1, \beta_2) = (-1, 2, 0.2)$, $\boldsymbol \alpha_{t,1}=(\alpha_{21,1}, \alpha_{31,1}, \alpha_{41,1})= (0.5, 0.7, 0.9)$, $\boldsymbol \lambda=(\lambda_1, \lambda_2)=(1, 1.05)$ and $b_{it} \sim N(0,\sigma^2_t)$, $\boldsymbol \sigma=(\sigma_2, \sigma_3, \sigma_4)=(0.66, 0.63, 0.60)$. $X_1$ was assumed to be a time independent variable. $X_2$ was taken as a response indicator variable for which while the first response took 1, the second one took 0. By the inclusion of response indicator as a covariate, we allowed bivariate responses to have different intercepts, i.e., while the intercept was $\beta_0+\beta_2 = -1+0.2 = - 0.8$ for the first response, it was $\beta_0 = -1$ for the second response. Additionally, the effect of $X_1$ was assumed to be shared across the responses, since interaction between $X_1$ and $X_2$ was not included in the model. Moreover, we assumed that the transition parameters were shared across responses, since $\boldsymbol Z_{itj}$ did not include response indicator variables, i.e., $Z_{itj}=[ \ 1 \ ]$.

We replicated the simulation study 200 times. Analysis of one simulated data (the last one) by PNMTREM(1) took 8.9 minutes on a PC with 4.00 GB RAM and 3.00 GHz processor. A simulated data set and the related procedure to analyze them can be found in the user manual of the {\bf pnmtrem} package.

The simulation results are displayed in Table \ref{tab:mtrem.sim.res}. Mean, bias, standard error of the parameter estimates (SE), mean of the standard error of the parameter estimates (meSE) and percentage coverage probabilities of the corresponding 95\% confidence intervals (CP\%) were calculated and reported. The marginal mean parameters of both baseline and $t \geq 2$ models were estimated very well. Put another way, the empirical biases of the parameter estimates were negligible: absolute biases lie between 0.005 (for $\beta_1^*$) and 0.014 (for $\beta_1$). The standard errors of the parameter estimates and the means of the standard error estimates were close to each other, e.g., these quantities were found identical for $\beta_2$ as 0.066. Moreover, the coverage probabilities were close to the nominal level 0.95, which indicate that the true values of the parameters were covered at the expected rate.

\begin{table}[t]
\centering
\caption{Simulation results for baseline and $t \geq 2$ models.}
\label{tab:mtrem.sim.res}
\scalebox{0.80}{
\fbox{%
\begin{tabular} {c c c c c c c}
\multicolumn{7}{c}{Baseline} \\ \hline 
Parameter     			& True   					    & Mean    & Bias    & SE    & meSE  & CP (\%)   \\ \hline
$\beta_0^*$   			& -1.000 					    & -1.011  & -0.010  & 0.130 & 0.131 & 95.5 \\ 
$\beta_1^*$   			& \  1.900 					    & \ 1.905 & \ 0.005 & 0.224 & 0.227 & 96.5 \\
\hline
\multicolumn{7}{c}{t $\geq$ 2} \\ \hline 
Parameter     			& True   					    & Mean    & Bias    &  SE   & meSE  & CP (\%) \\ \hline
$\beta_0$   			& -1.000 					    & -1.010  & -0.010  & 0.092 & 0.089 & 95.0 \\ 
$\beta_1$   			& \ 2.000 					    & \ 2.014 & \ 0.014 & 0.157 & 0.145 & 94.0 \\
$\beta_2$   			& \ 0.200					    & \ 0.206 & \ 0.006 & 0.066 & 0.066 & 96.0 \\
\end{tabular}}}
\end{table}

\section{Example: Iowa Youth and Families Project data set}

\subsection{Data}

The data set used to illustrate our model came from the Iowa Youth and Families Project (IFYP; Elder and Conger, 2000; Ilk, 2008). This project aimed to investigate the long term effects of the farm crisis, began in 1980's in America, on the well being of the family members living in the rural parts of the country. 451 families from eight rural parts of north central Iowa were selected. The focus was on 7th graders with two alive and biological parents and a sibling within 4 years of age. The study was started in 1989. Whereas it was conducted yearly until 1992, it was continued at 1994, 1995, 1997 and 1999. At each year, both the parents and the children of the aforementioned 451 families were surveyed. In the beginning of the study, the 7th graders were at average age of 12.7 years and 48\% of them were male (Ilk, 2008). Young people were followed during their adolescent period as well by this 11-year follow-up. 

The emotional statuses of young people were measured by three main distress variables, anxiety, hostility and depression (Table \ref{tab:iyfpvar2}). These variables were collected by a symptom check list, including nervousness, shakiness, an urge to break things and feeling low in energy etc, and dichotomized later according to whether having at least one of the distress symptoms (Ilk, 2008). It was observed that young people were highly distressed. For instance, almost 93\% of them reported at least one depression symptom at 1989 (Table \ref{tab:resp.perc}). It was also observed that young people tended to report higher depression compared to anxiety and hostility. Moreover the latter distress variables seemed to have close prevalences. A set of explanatory variables, which were thought to be related with these emotional variables, were also collected (Table \ref{tab:iyfpvar2}). These variables included gender, degree of negative life event experiences of the young people (such as having a close friend moved away permanently), financial cutbacks (such as moving to a cheaper residence) and negative economical event experiences of their families (such as changing job for a worse one). The main aim of collecting the family information was to measure the indirect effects of the farm crisis on the well-being of young people as well, e.g., due to harsh parenting. Among the explanatory variables, while gender was time-invariant, the others were time-varying. 

Transition model in the second level of PNMTREM required the use of equally spaced time points. We considered the first 4-year follow-up of the IYFP study, i.e., years 1989 to 1992, in our analyses, since this was a fully constrained portion of the whole data set (Ilk, 2008). Response and time indicator variables were included as additional explanatory variables, and dummy variables were created for all the categorical covariates (Table \ref{tab:iyfpvar2}). We coded the binary explanatory variables as 0 vs. 1 in our initial data analyses. However, an alternative coding, i.e., -1 vs. 1, was used due to convergence problems during model fittings with the initial analyses. The data set is available upon request from the authors. 

\begin{table}[t]
\centering
\caption{Variable list of IYFP used in PNMTREM(1).}
\label{tab:iyfpvar2}
\scalebox{0.90}{
\fbox{%
\begin{tabular} {l l}
Variable & Explanation\\ \hline
Responses & \\ 
anxiety & whether the young person had symptoms: 0=absence, 1=presence \\
hostility & whether the young person had symptoms: 0=absence, 1=presence \\
depression & whether the young person had symptoms: 0=absence, 1=presence \\ \hline
Covariates & \\ 
gender & gender of the young person: -1=male, 1=female \\
NLE1 & first indicator variable for negative life event experiences of young \\
 & people: 1=some, -1=none or many \\
NLE2 & second indicator variable for negative life event experiences of young \\
 & people: 1=many, -1=none or some \\
NEE & whether the household had any negative economical event: -1=no, 1=yes \\
cut1 & first indicator variable for financial cutback experiences of the household:\\
 & 1=between 1 and 5, -1= none or more than 5 \\
cut2 & second indicator variable for financial cutback experiences of the household:\\
 & 1=more than 5, -1= none or between 1 and 5 \\
resp1 & first response indicator variable: 1=hostility, -1=anxiety or depression \\
resp2 & second response indicator variable: 1=depression, -1=hostility or anxiety \\
time1 & first indicator variable for follow-up time: 1=1991, -1=1990 or 1992 \\
time2 & second indicator variable for follow-up time: 1=1992, -1=1990 or 1991 \\
\end{tabular}}}
\end{table}

\begin{table}[t]
\centering
\caption{Frequency table of the distress variables across years.}
\label{tab:resp.perc}
\scalebox{0.95}{
\fbox{%
\begin{tabular} {l c c c c }
	       & 1989         & 1990         & 1991         & 1992         \\ \hline
Anxiety    & 375 (83.2\%) & 347 (76.9\%) & 342 (75.8\%) & 327 (72.5\%) \\
Hostility  & 375 (83.2\%) & 350 (77.6\%) & 342 (75.8\%) & 328 (72.7\%) \\
Depression & 418 (92.7\%) & 385 (85.4\%) & 378 (83.8\%) & 386 (85.6\%) \\
\end{tabular}}}
\end{table}

\subsection{Relating data with the model}

PNMTREM(1) enables us to answer several questions on both the comparison of the sub-groups of young people and/or their families and on some specific young persons. Moreover, it permits drawing different statistical inferences for $t=1989$ and $t \geq 1990$ periods. For instance, we can compare the distress levels of males and females by the first levels of both baseline and $t \geq 2$ models. The inclusion of the interaction between gender and response indicator variables in the design matrices permits response specific comparison of the gender, i.e., comparison of anxiety, hostility and depression levels of males and females separately. We can measure the effect of the past year's distress status on the current ones by the second level of $t \geq 2$ model. The inclusion of the interaction between lag-1 responses and response indicator variables allows us to have response specific inferences about the transition probabilities. For instance, we can measure the relationship between the anxiety status of young people at 1990 and the ones at 1991. Furthermore, we can draw subject-specific inferences by using the last levels of the models. For instance, we can calculate the probability of being anxious for subject $223$ at year $1992$. Note that this probability is subject, time and response specific.

\subsection{Results}

We specifically built two different PNMTREM(1)'s. While the marginal regression parameters of these models were same, they differed in terms of separating the effects of the distress status histories on the current distress status for multiple responses. Put another way, the first model (Model 1 in Table \ref{tab:mtrem.iyfp.tgeq2}) included only ones in the design matrix $\boldsymbol Z_{itj}$, i.e., $Z_{itj}=[ \ 1 \ ]$. On the other hand, the second model (Model 2 in Table \ref{tab:mtrem.iyfp.tgeq2}) included response indicator variables in the design matrix $\boldsymbol Z_{itj}$, i.e., $\boldsymbol Z_{itj}=[ \ 1 \ \ resp1 \ \ resp2 \ ]$. Since the baseline models were same for Model 1 and Model 2, we presented only one baseline result in Table \ref{tab:mtrem.iyfp.baseline}. Results for $t \geq 2$ models are presented in Table \ref{tab:mtrem.iyfp.tgeq2}. In these tables, results of the generalized linear models (GLM) with \textit{probit} link are presented as well. Note that GLM ignored the within and multivariate response dependencies and the related results were actually used to start the Fisher-Scoring (F-S) algorithms. 

Since Model 1 and Model 2 are nested models, we can compare them by the likelihood ratio test (LRT). The corresponding maximized log-likelihoods were the summation of the ones for baseline and $t \geq 2$ models: -1236.78 $(=-210.78-1026)$ and -1234.49 $(=-210.78-1023.71)$ for Models 1 and 2, respectively. The LRT statistic for the comparison of these models was 4.58 $(=-2*(-1026-(-1023.71)))$ with a p-value of 0.60 which indicated that there was not enough evidence to conclude that Model 2 explained the IYFP data better compared to Model 1 with 95\% confidence level $(\chi^2_{6,0.95}=12.59)$. Therefore, throughout we only considered Model 1 while making interpretations about the parameter estimates. 

We checked the existence of possible multicollinearity problems via variance inflation factor (VIF). Results (not shown here) showed that there was no such problem in our models; the largest VIF was 1.17. Here, we also point out that Ilk and Daniels (2007) confirmed the exogeneity of the time-varying covariates in the IFYP data set. 

\begin{table}[t]
\centering
\caption{PNMTREM(1) and probit GLM results on IYFP data for $t=1989$. $H_0:\lambda^*_{hostility}=1$ and $H_0:\lambda^*_{depression}=1$; other parameters are tested for 0.}
\label{tab:mtrem.iyfp.baseline}
\scalebox{0.80}{
\fbox{%
\begin{tabular} {l r r r r r r r r}
\multicolumn{1}{c}{} 	   & \multicolumn{4}{c}{PNMTREM(1)} & \multicolumn{4}{c}{GLM}       		  \\ \hline 
Parameter     	     	   & Est.    &  SE  & Z         &   P      & Est.    & SE   & Z     & P 	  \\ \hline
$\beta_0^*$         	   & 1.33    & 0.07 & 18.82     & 0.00     & 1.27    & 0.07 & 19.65 & 0.00    \\
$\beta_{gender}^*$  	   & -0.09   & 0.06 & -1.41     & 0.16     & -0.10   & 0.06 & -1.75 & 0.08    \\ 
$\beta_{NLE1}^*$     	   & 0.20    & 0.12 & 1.61    	& 0.11     & 0.19    & 0.12 & 1.64  & 0.10    \\
$\beta_{NLE2}^*$  		   & 0.41    & 0.12 & 3.27    	& 0.00     & 0.39    & 0.12 & 3.24  & 0.00    \\
$\beta_{NEE}^*$     	   & 0.03    & 0.05 & 0.72    	& 0.47     & 0.03    & 0.05 & 0.67  & 0.50    \\ 
$\beta_{cut1}^*$     	   & 0.08    & 0.07 & 1.15    	& 0.25     & 0.07    & 0.06 & 1.11  & 0.27    \\
$\beta_{cut2}^*$    	   & -0.003  & 0.07 & -0.04   	& 0.97     & 0.01    & 0.06 & 0.12  & 0.91    \\
$\beta_{resp1}^*$	       & -0.001  & 0.06 & -0.02     & 0.99     & -0.0004 & 0.05 & -0.01 & 0.99    \\
$\beta_{resp2}^*$   	   & 0.29    & 0.07 & 4.17    	& 0.00     & 0.26    & 0.06 & 4.47  & 0.00    \\
$\beta_{gender*resp1}^*$ & -0.04   & 0.06 & -0.73   	& 0.47	   & -0.04   & 0.05 & -0.72 & 0.47    \\
$\beta_{gender*resp2}^*$ & -0.08   & 0.07 & -1.29   	& 0.20     & -0.09   & 0.06 & -1.47 & 0.14    \\ \hline
$\lambda_{hostility}^*$  & 1.10    & 0.79 & 0.12    	& 0.91 	   &         &      &       &         \\ 
$\lambda_{depression}^*$ & 1.04    & 0.71 & 0.05    	& 0.96 	   &         &      &       &         \\ 
$\mbox{log} (\sigma_1)$  & -0.41   & 0.41 &         	&      	   &         &      &       &         \\ \hline
$\mbox{Max. loglik.}$    & -210.78 &	    &		      	&		       & -511.98 &	    &  	  	&	   	  \\
\end{tabular}}}
\end{table}

Baseline results (Table \ref{tab:mtrem.iyfp.baseline}) indicated that only the intercept, one of the negative life event indicators (NLE2) and one of the response indicators (resp2) were significant at 95\% confidence level in 1989. The estimate of intercept ($\hat{\beta}_0^*=1.33$) indicated that young people had high probability of distress at 1989. Additionally, the estimate of the second response indicator variable ($\hat{\beta}_{resp2}^*=0.29$) indicated that young people were more likely to report depression compared to anxiety and hostility. Moreover, the insignificance of the first response indicator variable (p-value=0.99) indicated that there was not enough evidence towards differences in terms of reporting anxiety and hostility. These conclusions were indeed in agreement with our expectations, since the percentages of distress presences were fairly high for each response variables, and these percentages were higher for depression compared to anxiety and hostility (Table \ref{tab:resp.perc}). Moreover, it was found that young people who had many negative life events were more likely to be distressed ($\hat{\beta}_{NLE2}^*=0.41$). There was not enough evidence to say that the pairwise correlations between anxiety, hostility and depression were significantly different; corresponding p-values of $\lambda_{hostility}^*$ and $\lambda_{depression}^*$ were 0.91 and 0.96. The standard deviation of the random effects distribution was estimated as 0.66 $(=exp(-0.41))$ with a standard error of 0.27 ($=\sqrt{0.41^{2}*exp(-0.41*2)}$, by the delta method). This parameter was found highly significant with a p-value of 0.007. Of note, the calculation of this p-value was modified by following Molenberghs and Verbeke (2007), since the related hypothesis test introduced the equality of the parameter at its lower boundary. Although the marginal mean results of PNMTREM(1) and probit GLM seemed to be in agreement, the significant difference between the maximized log-likelihoods of these models (-210.78 and -511.98, respectively) indicated that fitting a marginalized random effects model explained the 1989 data better.

\begin{table}[t]
\centering
\caption{PNMTREM(1) and probit GLM results on IYFP data for $t \geq 1990$. $H_0:\lambda_{hostility}=1$ and $H_0:\lambda_{depression}=1$; other parameters are tested for 0.}
\label{tab:mtrem.iyfp.tgeq2}
\scalebox{0.80}{
\fbox{%
\begin{tabular} {l r r r r r r r r r r r r}
\multicolumn{1}{c}{}   & \multicolumn{8}{c}{PNMTREM(1)} & \multicolumn{4}{c}{GLM} \\ \cline{2-9}
\multicolumn{1}{c}{}   & \multicolumn{4}{c}{Model 1} & \multicolumn{4}{c}{Model 2} & \multicolumn{4}{c}{}              \\ \hline  
Parameter              & Est.   &  SE  & Z     &   P   & Est.   & SE   & Z     & P      & Est.    & SE   & Z     & P  \\ \hline
$\beta_0$              & 0.96   & 0.05 & 20.77 & 0.00  & 0.96   & 0.05 & 19.49 & 0.00   & 0.97    & 0.04 & 24.85 & 0.00 \\
$\beta_{gender}$       & 0.18   & 0.03 & 5.87  & 0.00  & 0.18   & 0.03 & 5.87  & 0.00   & 0.18    & 0.03 & 6.07  & 0.00 \\
$\beta_{NLE1}$         & 0.14   & 0.04 & 3.09  & 0.00  & 0.14   & 0.05 & 3.05  & 0.00   & 0.14    & 0.05 & 3.03  & 0.00 \\
$\beta_{NLE2}$         & 0.38   & 0.05 & 7.95  & 0.00  & 0.38   & 0.05 & 7.90  & 0.00   & 0.38    & 0.05 & 7.78  & 0.00 \\
$\beta_{NEE}$          & 0.08   & 0.03 & 3.08  & 0.00  & 0.08   & 0.03 & 3.03  & 0.00   & 0.07    & 0.02 & 2.98  & 0.00 \\
$\beta_{cut1}$         & 0.06   & 0.03 & 2.10  & 0.04  & 0.07   & 0.03 & 2.20  & 0.03   & 0.07    & 0.03 & 2.21  & 0.03 \\
$\beta_{cut2}$         & 0.02   & 0.03 & 0.72  & 0.47  & 0.02   & 0.03 & 0.73  & 0.47   & 0.03    & 0.03 & 0.80  & 0.42 \\
$\beta_{resp1}$        & 0.01   & 0.04 & 0.28  & 0.78  & 0.01   & 0.04 & 0.13  & 0.90   & 0.002   & 0.03 & 0.07  & 0.94 \\
$\beta_{resp2}$    	   & 0.22   & 0.04 & 5.21  & 0.00  & 0.22   & 0.05 & 4.66  & 0.00   & 0.21    & 0.04 & 5.90  & 0.00 \\
$\beta_{time1}$  	     & -0.07  & 0.04 & -1.75 & 0.08  & -0.08  & 0.05 & -1.75 & 0.08   & -0.05   & 0.04 & -1.36 & 0.18 \\
$\beta_{time2}$        & -0.09  & 0.05 & -1.96 & 0.05  & -0.09  & 0.05 & -1.88 & 0.06   & -0.06   & 0.04 & -1.80 & 0.07 \\
$\beta_{gender*resp1}$ & -0.01  & 0.03 & -0.18 & 0.86  & -0.01  & 0.03 & -0.20 & 0.84   & -0.0003 & 0.03 & -0.01 & 0.99 \\
$\beta_{gender*resp2}$ & 0.07   & 0.04 & 2.08  & 0.04  & 0.07   & 0.04 & 2.07  & 0.04   & 0.07    & 0.03 & 2.32  & 0.02 \\
$\beta_{resp1*time1}$  & -0.002 & 0.03 & -0.07 & 0.95  & -0.02  & 0.04 & -0.42 & 0.68   & -0.01   & 0.03 & -0.18 & 0.86 \\
$\beta_{resp1*time2}$  & 0.004  & 0.04 & 0.10  & 0.92  & 0.003  & 0.04 & 0.07  & 0.94   & -0.004  & 0.03 & -0.13 & 0.89 \\
$\beta_{resp2*time1}$  & -0.01  & 0.04 & -0.36 & 0.72  & -0.01  & 0.04 & -0.31 & 0.75   & -0.01   & 0.04 & -0.19 & 0.85 \\
$\beta_{resp2*time2}$  & 0.05   & 0.04 & 1.15  & 0.25  & 0.05   & 0.04 & 1.03  & 0.30   & 0.04    & 0.04 & 1.15  & 0.25 \\ \hline
$\alpha_{21,1}$ 	     & 0.76   & 0.11 & 6.62  & 0.00  & 0.75   & 0.17 & 4.50  & 0.00   & & & & \\
$\alpha_{22,1}$ 	     &        &      &       &  	   & 0.06   & 0.13 & 0.43  & 0.67   & & & & \\
$\alpha_{23,1}$ 	     &        &      &       &  	   & 0.11   & 0.16 & 0.70  & 0.48   & & & & \\
$\alpha_{31,1}$ 	     & 0.87   & 0.10 & 9.11  & 0.00  & 0.86   & 0.13 & 6.58  & 0.00   & & & & \\
$\alpha_{32,1}$ 	     &        &      &       &  	   & 0.08   & 0.11 & 0.74  & 0.46   & & & & \\
$\alpha_{33,1}$ 	     &   	    &      &       & 	     & 0.07   & 0.14 & 0.48  & 0.63   & & & & \\
$\alpha_{41,1}$ 	     & 0.90   & 0.10 & 9.53  & 0.00  & 0.86   & 0.12 & 7.03  & 0.00   & & & & \\
$\alpha_{42,1}$ 	     &  	    &      &       &  	   & -0.04  & 0.12 & -0.34 & 0.74   & & & & \\
$\alpha_{43,1}$ 	     &  	    &      &       &  	   & 0.12   & 0.13 & 0.93  & 0.35   & & & & \\ \hline
$\lambda_{hostility}$  & 1.03   & 0.37 & 0.60  & 0.94  & 0.99   & 0.36 & -0.02 & 0.99   & & & & \\
$\lambda_{depression}$ & 1.21   & 0.49 & 0.57  & 0.68  & 1.18   & 0.49 & 0.36  & 0.72   & & & & \\
$\mbox{log}(\sigma_2)$ & -0.48  & 0.25 &       &  	   & -0.47  & 0.26 &	     &	      & & & & \\
$\mbox{log}(\sigma_3)$ & -0.62  & 0.25 &       &  	   & -0.59  & 0.26 &  	   &	      & & & & \\
$\mbox{log}(\sigma_4)$ & -0.62  & 0.26 &       &  	   & -0.59  & 0.26 &       &        & & & & \\ \hline
$\mbox{Max. loglik}$   &  -1026.00      &   &  &       &  -1023.71      &    & & 	      & -1989.23 &  & & \\
\end{tabular}}}
\end{table}

For the later time points $(1990-1992)$, intercept, gender, both negative life event indicators (NLE1, NLE2), negative economical events experience (NEE), one of the cutbacks indicators (cut1), one of the response indicators (resp2), one of the time indicators (time2) and the interaction between gender and second response indicator (gender*resp2) were significant at 95\% confidence level (Table \ref{tab:mtrem.iyfp.tgeq2}). Similar to the baseline results, the estimate of the intercept indicated that young people had high probability of distress ($\hat{\beta}_{0}=0.96$). However, the distress probabilities tended to be lower than baseline ($\hat{\beta}_{0}^* > \hat{\beta}_{0}$). Females were more likely to be distressed compared to males ($\hat{\beta}_{gender}=0.18$). Moreover, they were more likely to be depressed ($\hat{\beta}_{gender*resp2}=0.07$) compared to them being anxious or hostile. Note that gender was found insignificant at 1989, and this result was supported by Ge et al. (2001, cited in Ilk, 2008) and Ilk (2008). Young people who experienced many negative life events and whose families experienced any negative economical events were found more likely to be distressed ($\hat{\beta}_{NLE1}=0.14$, $\hat{\beta}_{NLE2}=0.38$ and $\hat{\beta}_{NEE}=0.08$). Young people were more likely to be depressed compared to being anxious or hostile ($\hat{\beta}_{resp2}=0.22$). On the other hand, there was no significant difference between being anxious and hostile (p-value of $\beta_{resp1}=0.78$). While the distress levels were lower at 1992 compared to 1990 and 1991 ($\hat{\beta}_{time2}=-0.09$), there was no significant difference between 1990 and 1991 (p-value of $\beta_{resp1}=0.08$). However, the decrease in 1992 was not significantly different with respect to a specific response variable; p-values for ${\beta}_{resp1*time2}$ and ${\beta}_{resp2*time2}$ were found to be 0.92 and 0.25, respectively.

We can also interpret our {\em probit} marginal mean parameters as in the case of {\em logit} estimates, i.e., in terms of odds-ratios by using the JKB constant. This offers an approximate relationship between the {\em probit} and {\em logit} estimates, i.e., $\beta_{logit} \cong c*\beta_{probit}$ where $c=(15/16)(\pi/\sqrt{3})=1.700437$. For instance, the influence of the degree of negative life events on the probability of being distressed can be interpreted as follows: young people who experienced many negative life events were approximately 2.26 $(=exp(1.700437*((-1*0.14+ 1*0.38)-(1*0.14-1*0.38))))$ times more likely to be distressed compared to those with some negative life events, and individuals in the latter group were 1.60 $(=exp(1.700437*((1*0.14-1*0.38)-(-1*0.14-1*0.38))))$ times more likely to be distressed compared to those with no negative life events. 

The positive (and significant) transition parameter estimates indicated that young people who were distressed at year $t-1$ were more likely to be distressed at year $t$ compared to the ones who were not distressed at year $t-1$, i.e., $\hat{\alpha}_{21,1}=0.76$, $\hat{\alpha}_{31,1}=0.87$, $\hat{\alpha}_{41,1}=0.90$ with p-values $< 1 \times 10^{-10}$. Moreover, these transition parameter estimates were shared across anxiety, hostility and depression. As for the baseline model, there was not enough evidence to say that the pairwise correlations between anxiety, hostility and depression were significantly different; corresponding p-values were 0.94 and 0.68 for hostility and depression, respectively. The estimates of the standard deviations of the random effects distributions were found to be 0.62 $(=exp(-0.48))$, 0.54 $(=exp(-0.62))$ and 0.54 $(=exp(-0.62))$, respectively at 1990, 1991 and 1992. Related standard errors were 0.16, 0.14 and 0.14, respectively, and all of these parameters were found to be highly significant with p-values $<$ 0.0001. These results indicated that the individual variations were decreasing across time (recall that $\hat{\sigma}_1$=0.66) and close to each other at 1991 and 1992. Similar to the baseline results, $t \geq 2$ results indicated that PNMTREM(1) yielded a great improvement compared to GLM, which was apparent from the comparisons of the log-likelihoods, -1026 vs. -1989.23. 

Our PNMTREM(1) results for the IYFP data set mostly coincided with the ones reported by Ilk (2008). We observed that Model 1 and Model 2 produced equal or nearly same marginal regression parameter estimates, Z statistics and p-values. This is natural due to the fact that marginalized models are less sensitive to the misspecification of the dependence structures (Heagerty and Kurland, 2001). Moreover, Heagerty (2002) and Lee and Mercante (2010) proved that the parameters of the first and second levels of marginalized transition models (MTM) were orthogonal. Since the first and the second levels of PNMTREM are equivalent to MTM with \textit{probit} links, we expect the same property to hold for PNMTREM as well.

\begin{landscape}

\begin{figure}
\begin{center}$
\begin{array}{cc}
\includegraphics[width=3.4in]{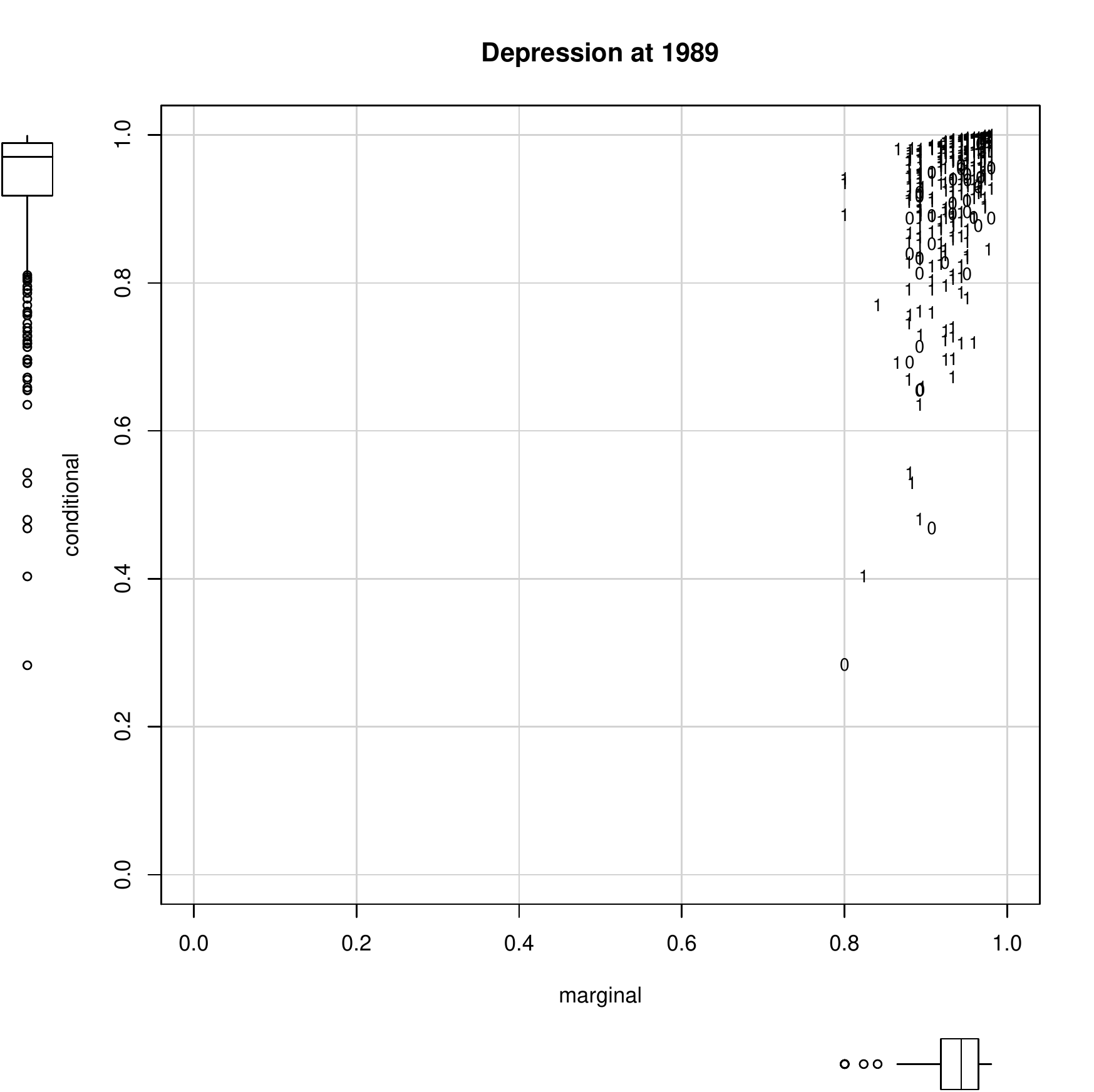} &
\includegraphics[width=3.4in]{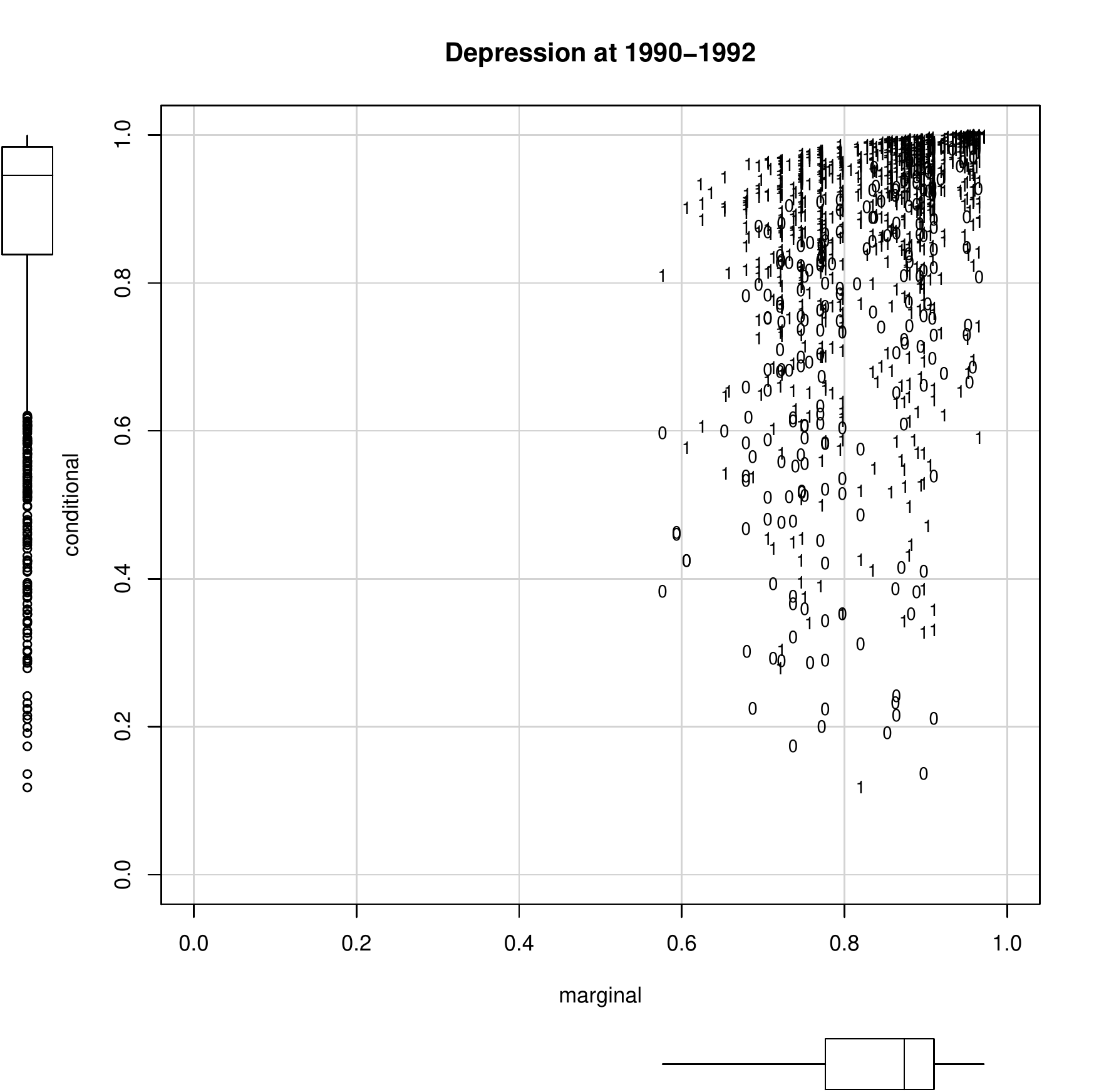}
\end{array}$
\end{center}
\caption{Scatter and box plots of marginal vs. conditional probabilities for response=depression at 1989 (left panel) and 1990-1992 (right panel).}
\label{fig:depr.cond.vs.marg}
\end{figure}

\end{landscape}

Up to here, we have drawn population-averaged inferences. Besides, we can draw individual-level inferences by using PNMTREM(1). To illustrate, we calculated the success probabilities regarding anxiety, hostility and depression of each person at each year by using the last levels of baseline and $t \geq 2$ models. In addition to these random effects probabilities, we calculated marginal probabilities for comparison purposes. These probabilities are summarized in Figure \ref{fig:depr.cond.vs.marg}. Due to page limits, we only included the figures for depression here; others could be found in the supplementary material. In these graphics, the observed values were labeled by 0 and 1 for absence and presence of a distress variable, respectively. Whereas the conditional probabilities ranged almost between the lower and upper probability bounds, the marginal counterparts ranged in a narrower interval. For instance, while the marginal probabilities of being depressed at the period of 1990-1992 took only the values in the interval of (0.576, 0.971), the conditional probabilities ranged between 0.118 and 0.999. This means that even the young people who had actually no depression for that period were assigned more probability of being depressed by the marginal models which would yield wrong decisions. This will be verified by two different accuracy measures at the end of this subsection. On the other hand, the conditional probabilities were spread widely and they yielded higher rates of correct decisions. For instance, in Figure \ref{fig:depr.cond.vs.marg}, the 0's (observing no depression for a young person) were associated with lower conditional probabilities. The associated box-plots reflected the location and scale information of these marginal and conditional probabilities as well. For instance, whereas the box-plot of the conditional probabilities reflected a spread distribution and many outlying probabilities, the marginal counterparts reflected a stacked and narrow distribution. Marginal models only rely on how well the covariates explain the variation of the responses and ignore the individual characteristics. Put another way, two young people with same covariates but different unobserved features would have the same probability of being depressed based on the results of marginal models. However, in random effects models these individual features are accounted by the random effects parameters in addition to the covariate effects. The reason that marginal probabilities were stacked in a narrower interval and tended to assign high probabilities to the cases in which distress variables were absent was most probably due to these facts. 

We built simple linear regression models considering the probit of the conditional probabilities, $\Phi^{-1}(P^r(Y_{itj}))$, as dependent variables and the probit of the marginal probabilities, $\Phi^{-1}(P^m(Y_{itj}))$, as independent ones to measure how much the variation in the responses were explained by the covariates. R-squares of these models are presented in Table \ref{tab:rsquared.marg.cond}. We observed that covariates in the IYFP data did not explain the individual characteristics well, since only up to 33\% of the individual variations were explained by the covariates.

Interactive graphics, for instance the ones obtained by GGobi software (Cook and Swayne, 2007), might help to identify interesting people. For instance, we detected a young person with ID=223 who was a female with some negative life event experiences, no negative economical event experiences and cutbacks between 1 and 5 (except in 1992 at which her family did not experience any cutbacks) and who actually never reported any distress at the period of 1989-1992 (Table \ref{tab:individual.prob}). For this person, whereas the marginal model (Marginal in Table \ref{tab:individual.prob}) indicated high probabilities of being distressed, conditional models (Conditional) indicated low probabilities. This means that the latter is more likely to yield correct inferences and the advantage of it is due to the estimation of individual characteristics. For instance, the Empirical Bayes estimate of $z_{223}$ was found to be $-2.45$. This indicates that this person was less likely to report distress compared to an average person. 

\begin{table}[t]
\centering
\caption{R-squares of the simple linear models which were constructed by considering $\hat{\Delta}_{i1j}^*+ \hat{\lambda_j}^* \hat{b}_{i1}$ and $\hat{\Delta}_{itj}^*+ \hat{\lambda_j} \hat{b}_{it}$ as dependent variables and $\boldsymbol X_{i1j} \hat{\boldsymbol \beta}^*$ and $\boldsymbol X_{itj} \hat{\boldsymbol \beta}$ as independent variables.}
\label{tab:rsquared.marg.cond}
\fbox{%
\begin{tabular} {c c c }
\multicolumn{1}{c}{Response} &  \multicolumn{1}{c}{1989} & \multicolumn{1}{c}{1990-1992} \\ \hline
Anxiety   		&   0.29 	 &  0.24       \\ 
Hostility  		&   0.26     &  0.20       \\
Depression 		&   0.31 	 &  0.33 	   \\
\end{tabular}}
\end{table}

\noindent Conditional probabilities can also be calculated by assuming that the person is an average person ($\mbox{Conditional}^*$), i.e., $b_{it}=0$ and $P^r(Y_{itj}=1|\boldsymbol X_{itj},y_{it-1j},b_{it}=0)=\Phi(\Delta_{itj}^*)$. These are still subject/time/response specific probabilities, since $\Delta_{itj}^*$ holds subject/time/response specific information. For instance, whereas at 1992 the probability of having anxiety for the young person with ID=223 was estimated as 0.64 by the marginal model, this probability was calculated as 0.08 by the conditional model. Moreover, the conditional probability assuming that the person was an average person was estimated as 0.46. However, as expected, the latter probabilities were not as successful as the conditional probabilities, yet they seemed to be better than the marginal probabilities.

\begin{table}[t]
\centering
\caption{Illustration of marginal and conditional probabilities for a specific person, ID=223. While Conditional corresponds to the random effects probabilities calculated by $\Phi(\hat{\Delta}_{itj}^*+ \hat{\lambda_j} \hat{b}_{it})$, $\mbox{Conditional}^*$ corresponds to random effects probabilities calculated by $\Phi(\hat{\Delta}_{itj}^*)$. The empirical Bayesian estimate of individual characteristics is: $\hat{z}_{223}=-2.45$.}
\label{tab:individual.prob}
\fbox{%
\scalebox{0.80}{
\begin{tabular} {c l c c c c c c c c}
Time & Response   & Gender & NLE  & NEE & Cutbacks     & Observed  &  Marginal  & Conditional & $\mbox{Conditional}^*$  \\ \hline
     & Anxiety    & Female & Some & No  & Betw. 1 \& 5 & Absence   &   0.82     &   0.30        & 0.87 \\			       
1989 & Hostility  & Female & Some & No  & Betw. 1 \& 5 & Absence   &   0.80     &   0.23        & 0.85 \\
     & Depression & Female & Some & No  & Betw. 1 \& 5 & Absence   &   0.91     &   0.47        & 0.95 \\ \hline
     & Anxiety    & Female & Some & No  & Betw. 1 \& 5 & Absence   &   0.78     &   0.09        & 0.56 \\
1990 & Hostility  & Female & Some & No  & Betw. 1 \& 5 & Absence   &   0.78     &   0.09        & 0.58 \\
     & Depression & Female & Some & No  & Betw. 1 \& 5 & Absence   &   0.90     &   0.14        & 0.77 \\ \hline
     & Anxiety    & Female & Some & No  & Betw. 1 \& 5 & Absence   &   0.74     &   0.14        & 0.59 \\
1991 & Hostility  & Female & Some & No  & Betw. 1 \& 5 & Absence   &   0.74     &   0.13        & 0.59 \\
     & Depression & Female & Some & No  & Betw. 1 \& 5 & Absence   &   0.86     &   0.22        & 0.79 \\ \hline
     & Anxiety    & Female & Some & No  & None 	 	 & Absence   &   0.64     &   0.08        & 0.46 \\
1992 & Hostility  & Female & Some & No  & None 	     & Absence   &   0.65     &   0.08        & 0.47 \\
     & Depression & Female & Some & No  & None  	     & Absence   &   0.85     &   0.19        & 0.77 \\
\end{tabular}}}
\end{table}

\begin{figure}[t]
\begin{center}
\includegraphics[width=5.6in,height=2.6in]{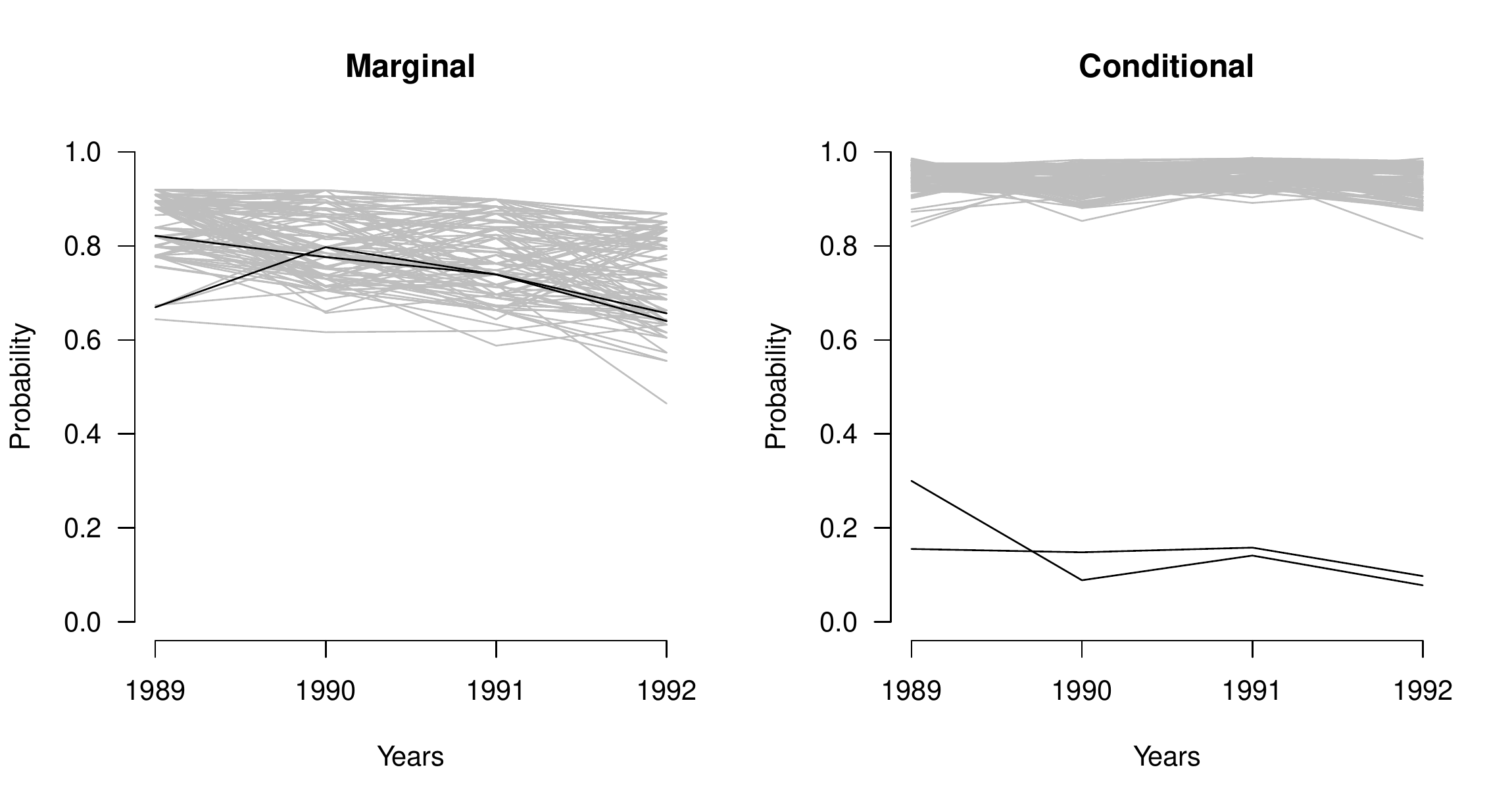} 
\end{center}
\caption{Spagetti plots of predicted marginal (left panel) and conditional (right panel) anxiety probabilities for subjects who reported the same answer for all  distress variables at all time points. While gray lines represent subjects who reported 1, the black lines represent subjects who reported 0.}
\label{fig:depr.stayers.all}
\end{figure}

Longitudinal binary data sets almost surely include subjects who constantly report absence (0) or presence (1) of a binary variable at all time points. For instance, in the IYFP data set, these subjects were the ones who reported absence or presence of anxiety, hostility and/or depression through all the follow-ups. Note that the subject with ID=223 constantly reported the absence of all distress variables. We identified such subjects in the IYFP data set in terms of three distress variables one-by-one and altogether. The counts and related percentages are given in Table \ref{tab:stayer.table}. There were considerable amount of subjects who reported the same answer through all study years. For instance, 29.7\% of the subjects reported 1 for all the three distress variables at all the time points. We calculated marginal and conditional probabilities for these subjects and summarized these probabilities in spagetti plots. Due to page limits, only the spagetti plot for the anxiety probability of subjects who reported the same answer for all three distress variables was included here (Figure \ref{fig:depr.stayers.all}). In this figure, while the gray lines represent the subjects who always reported 1, the black lines represent the ones who always reported 0. It was observed that the predictions were unsuccessful when a marginal model was used. With this model, the probability of distress was estimated high for all the young people who stayed at a single answer. In other words, the model was unable to distinguish the subjects who reported no stress over all years from the ones who reported stress through all follow-ups. On the other hand, our conditional probabilities were very successful at correctly assigning the success probabilities for these subjects; higher probabilities for subjects reporting 1 and lower probabilities for those who reported 0. Other spagetti plots indicated similar inferences (see the supplementary material).    

\begin{table}[t]
\centering
\caption{Frequency table for subjects who reported the same answer at all time points. ``All" stands for the subjects who reported the same answer for all distress variables.}
\label{tab:stayer.table}
\scalebox{0.95}{
\fbox{%
\begin{tabular} {l r r }
	          & Absence (0) & Presence (1)   \\ \hline
Anxiety     &  15 (3.3\%) & 215 (47.7\%) \\
Hostility   &  9  (2\%)   & 221 (49\%) \\
Depression  &  2  (0.4\%) & 288 (63.9\%) \\ 
All         &  2  (0.4\%) & 134 (29.7\%) \\
\end{tabular}}}
\end{table}

Finally, we considered two different accuracy measures to summarize the predicted probabilities. These measures are expected proportion of correct prediction (Herron, 1999) and area under the receiver operating characteristics curve (AUROC). Results (not shown here) showed that the inferences drawn from conditional models outperformed the ones drawn from the marginal models. This difference was apparent especially in terms of AUROC. For instance, while the value of the AUROC value for response=depression at 1990-1992 was found to be 0.684 for marginal models, this value was found to be 0.864 for the conditional models.

\section{Discussion and conclusion}

In this paper, we proposed a marginalized model for multivariate longitudinal binary data. The use of MLE and {\em probit} link facilitated the computations over BM and {\em logit} link. We proposed the use of implicit function theorem to solve the marginal constraint equations directly. To the best of our knowledge, this application was proposed for the first time here in marginalized structured models. An R package {\bf pnmtrem} was proposed to fit PNTREM(1), which was tested under different conditions with small studies. For the details and usage of the related function and examples, we refer the researchers to the package manual. The estimation of random effect coefficients within this package also allowed subject specific comparisons. We illustrated our model on the IYFP data set and discussed related parameter interpretations as well as subject specific inferences through the predicted probabilities.

A natural extension of our work here would be fitting higher order PNMTREM, PNMTEM(p) with $p > 1$. The variances of random effects could be modified by a subset of covariates, i.e., $\mbox{log}(\sigma_t)=\boldsymbol M_{itj} \ \boldsymbol \omega_{t}$ where $\boldsymbol M_{itj}$ is a possible subset of covariates and $\boldsymbol \omega_t$ are the related parameters. Also, the random effects might be assumed to have a multivariate normal distribution, i.e., $b_{it} \sim N(0, \boldsymbol D)$ where $\boldsymbol D$ is a $T \times T$ matrix. However, all of these possible extensions require intensive new derivations and implementations; hence, they are left as future work. \\ \\ \\
 
\noindent \textbf{\ Appendices} \\ 

\noindent \textbf{A. Linking second and third levels of PNMTREM(1)}\\
 
While linking second and third levels of the PNMTREM(1), we claim the following 

\begin{center}
$\int \Phi(\Delta_{itj}^*+\lambda_j b_{it}) f(b_{it}) db_{it}=\Phi \left(\frac {\Delta_{itj}^*} {\sqrt{1+\lambda_j^{2}\sigma_t^{2}}} \right)$
\end{center}

\noindent where $b_{it} \sim N(0,\sigma_{t}^2)$ and $b_{it}=z_i \sigma_t$, $z_i \sim N(0,1)$.\\

The related proof, which is modified from Griswold (2005), is given below. \\

Let $W_i \bot z_i$, where $W_i \sim N(0,1)$, then,

\begin{center}
$W_i/(\lambda_j \sigma_t) \sim N (0,(\lambda_j \sigma_t)^{-2})$\\
$W_i/(\lambda_j \sigma_t)-z_i \sim N (0,1+(\lambda_j \sigma_t)^{-2})$\\
$\frac {W_i/(\lambda_j \sigma_t)-z_i} {\sqrt{1+(\lambda_j \sigma_t)^{-2}}} \sim N(0,1) $
\end{center}

\noindent and

\begin{align*}
\int \Phi(\Delta_{itj}^*+\lambda_j b_{it}) f(b_{it}) db_{it} &=\int_{-\infty}^{+\infty} \Phi(\Delta_{itj}^*+\lambda_j z_i \sigma_t) \phi(z_i) dz_i\\
&= \int_{-\infty}^{+\infty} P(W_i \leq \Delta_{itj}^*+\lambda_j z_i \sigma_t) \phi(z_i) dz_i \\
&= \int_{-\infty}^{+\infty} P\left(\frac {W_i/(\lambda_j \sigma_t)-z_i} {\sqrt{1+(\lambda_j \sigma_t)^{-2}}} \leq \frac{\Delta_{itj}^*/ (\lambda_j \sigma_t)} {\sqrt{1+(\lambda_j \sigma_t)^{-2}}}\right) \phi(z_i) dz_i \\
&= P\left(\frac {W_i/(\lambda_j \sigma_t)-z_i} {\sqrt{1+(\lambda_j \sigma_t)^{-2}}} \leq \frac{\Delta_{itj}^*/ (\lambda_j \sigma_t)} {\sqrt{1+(\lambda_j \sigma_t)^{-2}}}\right)
= \Phi \left(\frac {\Delta_{itj}^*} {\sqrt{1+(\lambda_j \sigma_t)^{2}}} \right)
\end{align*}
 
\noindent \textbf{B. Details of first partial derivatives}\\ 
 
\noindent \textbf{B.1 Baseline model} \\
 
The derivatives of $d_{i1jq}$ with respect to $\boldsymbol \theta_1=(\boldsymbol \beta^*, \boldsymbol \lambda^*, c_1)$ with $\boldsymbol \lambda^*=(\lambda_2^*, \ldots, \lambda_k^*)$ are given below.

\small
\begin{align*}
\frac{\partial d_{i1jq}}{\partial \boldsymbol \beta^*}&=\sqrt{1+{\lambda_j^*}^2e^{2c_1}} (\boldsymbol X_{i1j}) \nonumber \\
\frac{\partial d_{i1jq}}{\partial \lambda_j^*}&=({1+{\lambda_j^*}^2e^{2c_1}})^{-1/2} \lambda_j^* e^{2c_1} (\boldsymbol X_{i1j} \boldsymbol \beta^*) + e^{c_1} \sqrt{2} z_q \nonumber \\
\frac{\partial d_{i1jq}}{\partial c_1}&=({1+{\lambda_j^*}^2e^{2c_1}})^{-1/2} {\lambda_j^*}^2 e^{2c_1} (\boldsymbol X_{i1j} \boldsymbol \beta^*) + \lambda^*_j e^{c_1} \sqrt{2} z_q
\end{align*} 
\normalsize 

\noindent \textbf{B.2 $t \geq 2$ model} \\
 
The derivatives of $d_{itjq}$ with respect to $\boldsymbol \theta_2=(\boldsymbol \beta, \boldsymbol \alpha_{t,1}, \boldsymbol \lambda, \boldsymbol c)$ with $\boldsymbol \lambda= (\lambda_2, \ldots, \lambda_k)$ and $\boldsymbol c= (c_2, \ldots, c_T)$ are given below.
\small
\begin{align*}
\label{eq:parderivtgeq2}
\frac{\partial d_{itjq}}{\partial \boldsymbol \beta}&=\sqrt{1+{\lambda_j}^2e^{2c_t}} (\boldsymbol A_{itj}) \nonumber\\
\frac{\partial d_{itjq}}{\partial \boldsymbol \alpha_{t,1}}&=\sqrt{1+{\lambda_j}^2e^{2c_t}} (\boldsymbol B_{itj}+\boldsymbol Z_{itj}y_{it-1j}) \nonumber\\
\frac{\partial d_{itjq}}{\partial \lambda_j}&=({1+{\lambda_j}^2e^{2c_t}})^{-1/2} \lambda_j e^{2c_t} \left(-(\boldsymbol A_{itj}\boldsymbol \beta_0+\boldsymbol B_{itj} \boldsymbol \alpha_{t,10})+\boldsymbol A_{itj} \boldsymbol \beta+ \boldsymbol \alpha_{t,1}(\boldsymbol B_{itj}+\boldsymbol Z_{itj}y_{it-1j})\right) + e^{c_t} \sqrt{2} z_q\nonumber\\
\frac{\partial d_{itjq}}{\partial c_t}&=({1+{\lambda_j}^2e^{2c_t}})^{-1/2} {\lambda_j}^2 e^{2c_t} \left(-(\boldsymbol A_{itj} \boldsymbol \beta_0+ \boldsymbol B_{itj} \boldsymbol \alpha_{t,10})+\boldsymbol A_{itj} \boldsymbol \beta+\boldsymbol \alpha_{t,1}(\boldsymbol B_{itj}+\boldsymbol Z_{itj}y_{it-1j})\right) + \lambda_j e^{c_t} \sqrt{2} z_q
\end{align*}
\normalsize 

\noindent where 
\begin{align*}
\boldsymbol A_{itj}=- \frac{\frac{\partial F}{\partial \boldsymbol \beta} \Big |_{(\boldsymbol \beta_0, \boldsymbol \alpha_{t,10},\Delta_{itj0})}} {\frac{\partial F}{\partial \Delta_{itj}} \Big |_{(\boldsymbol \beta_0,\boldsymbol \alpha_{t,10},\Delta_{itj0})}},
\; \; \boldsymbol B_{itj}=-\frac{\frac{\partial F}{\partial \boldsymbol \alpha_{t,1}} \Big |_{(\boldsymbol \beta_0,\boldsymbol \alpha_{t,10},\Delta_{itj0})}} {\frac{\partial F}{\partial \Delta_{itj}} \Big |_{(\boldsymbol \beta_0,\boldsymbol \alpha_{t,10},\Delta_{itj0})}}
\end{align*}

\end{document}